\documentclass[final,3p]{CSP}
\usepackage{amssymb}
\usepackage{amsmath, bm, array, booktabs, enumitem, makecell}
\usepackage{changepage, caption, subcaption, pdflscape}
\usepackage[online]{threeparttable}
\begin{document}

\begin{frontmatter}

\title{Group LASSO Variable Selection Method for Treatment Effect Generalization} %% need edits

\author[UMN]{Chuyu Deng}
\author[UNR]{Brandon Koch}
\author[UMN]{David M. Vock}
\author[UMN]{Joseph S. Koopmeiners}

\address[UMN]{University of Minnesota, Minneapolis}
\address[UNR]{The Ohio State University, Columbus}


\begin{abstract}\rm
\begin{adjustwidth}{2cm}{2cm}{\itshape\textbf{Abstract:}} 
Often in public health, we are interested in the treatment effect of an intervention on a population that is systemically different from the experimental population the intervention was originally evaluated in. When treatment effect heterogeneity is present in a randomized controlled trial, generalizing the treatment effect from this experimental population to a target population of interest is a complex problem; it requires the characterization of both the treatment effect heterogeneity and the baseline covariate mismatch between the two populations. Despite the importance of this problem, the literature for variable selection in this context is limited. In this paper, we present a Group LASSO-based approach to variable selection in the context of treatment effect generalization, with an application to generalize the treatment effect of very low nicotine content cigarettes to the overall U.S. smoking population. 
\end{adjustwidth}
\end{abstract}

\end{frontmatter}

\section{Introduction}

% External validity
Randomized controlled trials (RCTs) have long been considered the gold standard for estimating causal treatment effects, and much attention has been devoted to best practices for ensuring their internal validity \cite{Shadish, Weisberg}. However, experimental samples are often not representative of the target population of interest, and internal validity alone is not enough to answer many questions in public health \cite{Gheorghe, Allcott, Deaton}.
For many applications, we want to generalize the results of a RCT from a smaller experimental sample to a broader target population \cite{Shadish}, and the generalization of causal treatment effects when the experimental sample is not representative of a target population is an area of interest for researchers and policymakers. In particular, our motivating application is the generalization of the treatment effect of very low nicotine content cigarettes on cigarettes smoked per day from an experimental sample collected in a RCT to the overall U.S. smoking population. The question of generalization is of great importance in this application, where broad policy changes based on the results of the clinical trial would affect all U.S. smokers, but the experimental sample is not representative of our target population. 

% 2 major hurdles for generalization 
Whether or not the treatment effect estimated from an experimental sample can be generalized to a target population depends on two major statistical considerations: treatment effect heterogeneity and baseline covariate mismatch between the experimental sample and target populations. Treatment effect heterogeneity (TEH) refers to the phenomenon in which the same treatment can affect different people in disparate ways. If a treatment affected everyone in the population identically, the treatment effect estimated in the experimental sample would be directly applicable to the target population, and treatment effect generalization would not be necessary. 
Alternately, the treatment effect may vary throughout the sample as a function of one or multiple covariates, resulting in subgroup-specific treatment effects. In this case, understanding how the treatment effect varies across the sample is a crucial step in generalizing a study's results. 
Baseline covariate mismatch between the experimental sample and target populations refers to the situation in which the experimental sample is not a representative sample of the target population. If there is no covariate mismatch between the experimental sample and target population, the treatment effect estimated in the experimental sample would be directly applicable to the target population. In the case of treatment effect heterogeneity and covariate mismatch, both the TEH and covariate mismatch must be accounted for in order to generalize the treatment effect to the target population. 

% current methods for generalization
There has been an increasing interest in the development of statistical methods that address the problem of treatment effect generalization. \cite{Tipton2018}. 
Stuart et al.\@ proposed the use of three different propensity score methods (here the propensity is for inclusion in the experimental sample also referred to as the selection model) -- inverse probability weighting (IPW), full matching, and subclassification -- to evaluate if the experimental sample and the target population are similar enough for results to be generalized \cite{Stuart2011}. The propensity scores can also be used to re-weight covariates or individuals in the study to resemble the target population. 
Tipton et al.\@ developed a similar propensity score-based subclassification estimator \cite{Tipton2013}. 
Kern et al.\@ proposed an outcome regression modeling-based method using Bayesian additive regression trees (BART) \cite{Kern}. This method first fits the BART model for the outcome using a MCMC algorithm to the experimental sample. Then, the causal effect for each person in the target population is calculated from draws of the posterior distributions of the counterfactual outcomes. The average of these individual causal effects is taken as the average treatment effect of the target population. 
Finally, Chan et al.\@ proposed a bounding approach that gives an interval estimate instead of a point estimate of the average treatment effect in the target population \cite{Chan}. These interval estimates are a function of the proportion of overlap between the experimental and target populations, the treatment effect in the experimental sample, and assumptions regarding the minimum and maximum possible treatment impact. 

% VS problem with current generalization methods
In practice, when generalizing the results of a RCT to a target population, there may be many covariates available, but no certainty as to which covariates should be included in an outcome or propensity score model for treatment effect generalization. However, no paper to date has evaluated in detail if and how different subsets of covariates can impact the model performance of generalization methods in the absence of appropriate variable selection. 
In contrast, variable selection for potential confounders, which is central to estimating causal effects, has been studied extensively in the literature. 
These variable selection methods include a family of approaches based on model averaging such as 
Bayesian adjusting for confounding (BAC) \cite{Wang2015}, model averaged double robust (MA-DR) estimators \cite{Cefalu}, and Bayesian methods for propensity score variable selection and model averaging \cite{Zigler}. 
Another family of methods based on penalized regression include
an outcome adaptive lasso method for the propensity score model \cite{Shortreed}, 
a penalized credible region method (WR) \cite{Wilson}, 
and an adaptive group lasso approach to select covariates simultaneously in the outcome and covariate models \cite{Koch}. 
However, these existing variable selection methods are not able to concurrently address the problems of TEH and baseline covariate mismatch that are present in the case of generalization. We show in Section \ref{toyex} that when generalizing between two mismatched populations, it is important to adjust for baseline covariates associated with the treatment effect and detrimental to add in irrelevant covariates. Thus, variable selection for the purpose of generalizing the treatment effect from one population to another is an important concern. For the generalization of a treatment effect, we want to find a set of covariates that are related to the treatment effect and differentially distributed in the two populations, which requires a different set of covariates than those found with confounder variable selection methods. 
Recent papers have shown that under certain assumptions, by adjusting for a separating set, a group of covariates that affect both the sampling mechanism and treatment effect heterogeneity, the TATE can be identified \cite{Tipton2013, Kern, Egami}. 
Egami et al.\@ proposed a method that selects the separating set using a Markov random field to encode conditional independence relationships among observed covariates and existing domain knowledge. The TATE is then estimated with the separating set using an IPW estimator \cite{Egami}. This method allows the use of a sparser target dataset instead of requiring rich covariate data in both the experimental sample and target datasets. However, this method is not entirely data-driven as the sampling set, the set of variables that determines the sampling mechanism of the experimental sample, must be prespecified using domain knowledge. 

% sell our method
In this paper, we introduce GLAVeS (Group LAsso Variable Selection), a treatment effect generalization method that performs simultaneous variable selection for the selection and outcome models using a modified adaptive group lasso approach. The outcome model is then used to estimate the TATE. This method is able to capture covariates that are weak treatment interactions by supplementing the penalization of the treatment interaction terms with the probability of selection into the target population for each covariate. This allows GLAVeS to keep covariates that are weak treatment interactions that standard variable selection methods, which only considers the association of covariates with the outcome, would miss. %do we want to talk about OLSGlid?

% paper structure
We first demonstrate the necessity of proper variable selection in the context of treatment effect generalization in Section \ref{toyex}. In Section \ref{Gmethods}, we propose a novel data-driven method, GLAVeS, for simultaneously conducting variable selection and estimating the TATE. A simulation study is conducted in Section \ref{Gsims} to compare the performance of GLAVeS against other treatment effect generalization methods. We then apply our method to the problem of generalizing the effect of very low nicotine content (VLNC) cigarettes to the United States smoking population using the Population Assessment of Tobacco and Health (PATH) Study, a nationally representative survey of smokers in the United States. We conclude with a discussion in Section \ref{disc}. 

%%%%%%%%%%%%%%%%%%%%% MOTIVATING EXAMPLE %%%%%%%%%%%%%%%%%%%%
\section{Impact of Variable Selection for Generalization}
\label{toyex}

\subsection{Notation and Assumptions}
\label{notation}
In this paper, we will use the Sample Average Treatment Effect (SATE) and Target Average Treatment Effect (TATE) notation and assumptions outlined in Kern et al.\@ \cite{Kern}. We assume that the data analyst has two data sources: an experimental sample which has data on treatment, covariates, and outcome, and a representative sample from the target population which has at least covariate information. 
The experimental sample data consists of a continuous outcome $Y_i^{obs}$, a binary treatment assignment $A_i$ where $A_i=1$ denotes treatment and $A_i=0$ denotes control, and baseline covariates $\boldsymbol{X_{i}}=(X_{i1},..., X_{ip})$ for subjects $i=1,...,n$. Let $Y_i(a)$ be the potential outcome of an individual $i$ in the experimental data if they were to receive treatment $a$ for $a \in \{0,1\}$. 
The sample from the target population will only have the baseline covariates $\boldsymbol{X^*_{ij}}$ where $j=1,...,p$, for subjects $i=1,...,m$. 
Let $Y^*_i(a)$ be the potential outcome of a random individual $i$ in the target data if they were to receive treatment $a$ for $a \in \{0,1\}$. 
For the remainder of the article, we suppress $i$ in our notation except when necessary. 

Certain standard causal assumptions are required here including the stable unit treatment value assumption (SUTVA), positivity, and no unmeasured confounders ($A\perp\{Y(1), Y(0)\} |X$). No unmeasured confounding is guaranteed in randomized studies which is what we consider in this chapter. These assumptions have been discussed and established in previous works on generalizability such as in Stuart et al.\@ \cite{Stuart2011}.  Additionally, we will also need to make an assumption that the selection mechanism be strongly ignorable given the baseline covariate information. This assumption means that for an individual $i$, the distribution of their treatment effect $Y_i(1)-Y_i(0)$ is independent of belonging to the experimental or target population given their baseline covariates $X_i$, which implies along with the other causal assumptions stated above, that $E(Y|A, \boldsymbol{X}) = E(Y^*|A, \boldsymbol{X})$.

We define the SATE as the causal estimand that averages over individual-level causal effects in the experimental population:
\begin{align*}
        \Delta &= E(Y(1) - Y(0)) \\
        &= \int \Big( E(Y(1) | \boldsymbol{X}) - E(Y(0)|  \boldsymbol{X})\Big)f(\boldsymbol{X})  d\boldsymbol{X} \\
 &= \int \Big( E(Y|A=1, \boldsymbol{X}) - E(Y|A=0,  \boldsymbol{X})\Big)f(\boldsymbol{X})  d\boldsymbol{X} 
\end{align*}
where $f(\boldsymbol{X})$ is the density function of the covariates in the experimental population. The second equality holds based on the causal assumptions made prior. 
We can posit models for $E(Y|A=1, \boldsymbol{X}, \boldsymbol{\beta}_{A=1})$ and $E(Y|A=0, \boldsymbol{X}, \boldsymbol{\beta}_{A=0})$ which may depend on a vector of parameters $\boldsymbol{\beta}_{A=1}$ and $\boldsymbol{\beta}_{A=0}$ using the individuals assigned treatment and control, respectively, in the experimental sample. We can estimate these parameters using standard methods (e.g. ML estimators). 
If we use the empirical distribution from the experimental sample to estimate the $f(\boldsymbol{X})$ (i.e. $\hat f(\boldsymbol{X})$) then the plug-in estimator for $\Delta$ is then: 
\begin{align*}
        \hat \Delta &= \int \Big( 
            E(Y|A=1, \boldsymbol{X}, \hat{\boldsymbol{\beta}}_{A=1}) - 
            E(Y|A=0, \boldsymbol{X}, \hat{\boldsymbol{\beta}}_{A=0}) \Big) 
            \hat f(\boldsymbol{X}) d \boldsymbol{X} \\
        &= \frac{1}{n} \sum^{n}_{i=1} 
            E(Y|A=1, X_i, \hat{\boldsymbol{\beta}}_{A=1}) -
            E(Y|A=0, X_i, \hat{\boldsymbol{\beta}}_{A=0}) 
\end{align*}
We define the TATE as the causal estimand that averages over individual-level causal effects in the target population:
\begin{align*}
    \Delta^* &= \int \Big( E(Y^*|A=1, \boldsymbol{X}^*) - E(Y^*|A=0, \boldsymbol{X}^*)\Big) f(\boldsymbol{X}^*) d \boldsymbol{X}^* \\
    &= \int \Big( E(Y|A=1, \boldsymbol{X}^*) - E(Y|A=0, \boldsymbol{X}^*)\Big) f(\boldsymbol{X}^*) d \boldsymbol{X}^*
\end{align*}

What we define as the TATE is sometimes referred to in other work on generalization as the Population Average Treatment Effect (PATE). In our opinion, the TATE terminology (“target”) is more intuitive and flexible since it acknowledges that an experimental estimate may be generalized to several target populations.
To obtain the TATE, $E(Y^*|A=1, \boldsymbol{X}^*)$ and $E(Y^*|A=0, \boldsymbol{X}^*)$ can be estimated by using the estimates for $E(Y|A=0, \boldsymbol{X})$ and $E(Y|A=1, \boldsymbol{X})$ from the experimental population under the assumption of strong ignorability of the selection mechanism. Thus the plug-in estimator for $\Delta^*$ is: 
\begin{align}
        \hat \Delta &= \int \Big( 
            E(Y|A=1, \boldsymbol{X}^*, \hat{\boldsymbol{\beta}}_{A=1}) - 
            E(Y|A=0, \boldsymbol{X}^*, \hat{\boldsymbol{\beta}}_{A=0}) \Big) 
            \hat f(\boldsymbol{X}^*) d \boldsymbol{X}^* \nonumber \\ 
        &= \frac{1}{m} \sum^{m}_{i=1} 
            E(Y|A=1, X_i^*, \hat{\boldsymbol{\beta}}_{A=1}) -
            E(Y|A=0, X_i^*, \hat{\boldsymbol{\beta}}_{A=0}) 
\label{tate.eq}
\end{align}

\subsection{Simulation Study}

% when have TEH and mismatch, what variables are important? 
From the existing literature on treatment effect generalization, we know that the TATE can be identified using a group of covariates that affect both the sampling mechanism and TEH \cite{Tipton2013, Kern, Egami}. However, while these results specify the minimal set of covariates that must be included for the TATE to be identifiable, they do not provide intuition as to the effect of proper variable selection on bias and efficiency. 

We conducted a straightforward simulation study to illustrate the importance of adequate variable selection in a model for treatment effect generalization, especially when the experimental sample and target populations have differentially distributed covariate distributions. 
For simplicity, only one covariate was used in this simulation, but the results can be extrapolated to multiple covariates. 

% data generation 
We simulated two scenarios: 
1. The experimental sample and target populations have the same covariate distribution
2. The experimental sample and target populations have differentially distributed covariate distributions. 
In both scenarios, the experimental dataset consists of 300 observations, containing a binary treatment variable ($A$), a covariate ($X$), an outcome variable ($Y$), and the sample from the target population consists of 900 observations with a single covariate ($X^*$). 
Under the scenario where the experimental and target populations have the same covariate distribution, both $X$ and $X^*$ are simulated from a $N(0,1)$. 
Under the scenario where the experimental and target populations are differentially distributed, $X$ is simulated from a $N(0,1)$, and $X^*$ is simulated from a $N(1,1)$. 

The outcome variable in the experimental sample dataset was generated to match the 3 main covariate types that we are interested in: 
1. Covariate is unrelated to outcome variable $(E(Y|A,X) = A)$
2. Covariate is related to outcome variable $(E(Y|A,X) = A+X)$
3. Covariate is related to outcome variable and interacts with the treatment $(E(Y|A,X) = A+X+AX)$.
This creates 6 different data generation scenarios in total.

\begin{figure}[!ht]
\centering
\begin{threeparttable}
\begin{tabular}{cc}
\toprule
\textbf{Experimental Sample} & \textbf{Target Population} \\
\midrule
\makecell[lt]{
$\begin{aligned}[t]
X_i \sim& \begin{cases} 
N(0,1), & \text{for ``Same Covariate Distribution" case} \\
N(1,1), & \text{for ``Different Covariate Distribution" case} \end{cases} \\
A_i \sim& Ber(0.5)\\
Y_i =& \begin{cases} 
\begin{bmatrix}A & X & AX \end{bmatrix} \begin{bmatrix}1 & 0 & 0 \end{bmatrix}^T + \varepsilon_i, & \text{for ``} Y \text{ unrelated to } X \text{" case}\\
\begin{bmatrix}A & X & AX \end{bmatrix} \begin{bmatrix}1 & 1 & 0 \end{bmatrix}^T + \varepsilon_i, & \text{for ``} Y \text{ related to } X \text{" case}\\
\begin{bmatrix}A & X & AX \end{bmatrix} \begin{bmatrix}1 & 1 & 1 \end{bmatrix}^T + \varepsilon_i, & \text{for ``} Y \text{ related to } AX \text{" case} \end{cases}\\
&\text{where }\varepsilon_i \sim N(0,1.5)  
\end{aligned}$ }
& 
\makecell[tl]{
$\begin{aligned}[t]
X^*_i &\sim N(0,1)\\
\end{aligned}$ }\\
\bottomrule
\end{tabular}
\end{threeparttable}
\caption{Data generation for the 6 data scenarios of the simulation study.}
\label{toyDG}
\end{figure}

%the models
We then evaluated the performance of the 3 following linear models which we posited for the outcome: 
\begin{align*}
\mathbb{E}(Y|A,X) &=\alpha_0+\alpha_1A \\
\mathbb{E}(Y|A,X) &=\beta_0+\beta_1A+\beta_2X \\
\mathbb{E}(Y|A,X) &=\gamma_0+\gamma_1A+\gamma_3X+\gamma_4AX
\end{align*}
We fit each model using the experimental sample dataset, and then estimated the outcomes under each scenario using the corresponding sample from the target population to calculate the TATE. 
This procedure was repeated 1000 times for each data generation scenario and model, and the resulting MSEs of the estimated TATE are shown in Figure \ref{msetab}. 

\begin{table} [bt]
\centering
\includegraphics[width=0.9\textwidth]{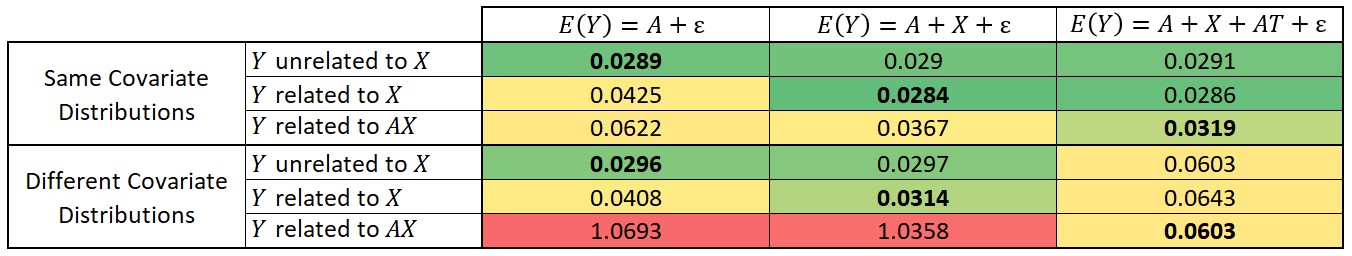}
\caption{MSEs for the 6 data scenarios under 3 different linear models. The columns are the assumed models, and the rows are the true data generating models. }
\label{msetab}
\end{table}

Under the scenarios where the experimental sample and target populations are not differentially distributed, we see that as expected, the assumed models that match each data generation model had the lowest MSEs for the estimated TATE. We also see that having extraneous terms (i.e., covariates with cofficients truly equal to zero) in these models do not seem to substantially impact their performances, as long as the correct terms are present. 
Under the scenarios where the experimental sample and target populations are differentially distributed, the models that match each data generation model resulted in the lowest MSEs. In this case, we see that missing a treatment interaction of a differentially distributed covariate can be detrimental to estimating the TATE correctly, which matches our intuition and the results from Egami et al.\@ \cite{Egami}. 
However, including extraneous interaction terms in the model has an adverse impact on MSE, nearly doubling the MSE when included unnecessarily, highlighting the importance of targeted variable selection. 

%Need to concretely argue why this is relevant for more realistic scenarios
Although this simulation study was designed with only 1 covariate, it effectively showcases the penalties to the MSE that would occur under all the possible inclusion/exclusion patterns for a standard linear model with treatment interaction. We argue that in more realistic scenarios where covariates could be ``drawn" from any one of these cases we outlined, the penalties to the MSE shown would still be applicable. 

%The penalties to the MSE shown here demonstrates the need for variable selection in the context of treatment effect generalization when the sample and target populations have differential covariate distributions, and motivates the novel method presented in this paper. 

%%%%%%%%%%%%%%% METHODS %%%%%%%%%%%%%%%%%%%%
\section{Methods}
\label{Gmethods}

\subsection{GLAVeS} 
\label{glid}
The goal of GLAVeS is to capture covariates that are weak treatment interactions that would not be selected by standard variable selection methods that only consider the predictive performance of covariates to the outcome. We achieve this by incorporating the association of each covariate with the probability of selection into the experimental sample in the penalization of interaction terms in the outcome model. This allows us to keep covariates that are weak treatment interactions but strongly differentially distributed between the experimental sample and target populations. 

GLAVeS is similar to GLiDeR \cite{Koch}, which uses a modified adaptive group lasso approach to perform simultaneous variable selection for the treatment and outcome models, and estimation of the average treatment effect. GLiDeR is able to capture covariates that might be weakly associated with the outcome, but strongly associated with the treatment assignment. In our case, since it is detrimental to omit covariates that are associated with the treatment effect (i.e., covariates which interact with treatment), while also associated with selection, we use a similar framework to capture covariates that are weakly associated with the treatment effect, but strongly mismatched between the experimental sample and target populations. 
As in the original group lasso, grouped covariates are forced to have either all zero or all nonzero coefficients. GLiDeR modifies this setup so that instead of grouping distinct covariates within the same regression model, it groups the same covariates across a treatment model and an outcome model. We modify this setup such that a selection model replaces the treatment model, and the outcome model contains both main effects and interactions with treatment:
\begin{align}
    \text{Outcome model:} \nonumber\\
    E(Y|A,\boldsymbol{X}) &= \sum^p_{j=1}\beta_jX_{j} + \sum^p_{j=1}\beta_{p+j}X_{j}A + \beta_{2p+1} + \beta_{2p+2}A  \\
    \text{Selection model:} \nonumber\\
    logit(P(S=1|\boldsymbol{X}^C)) &= \sum^p_{j=1}\gamma_jX^C_{j} + \gamma_{p+1} \label{selmodel}
\end{align}

Where $\boldsymbol{X}^C = \boldsymbol{X}$ if a subject is from the experimental sample and $\boldsymbol{X}^C =\boldsymbol{X}^*$ if a subject is from the target sample. 
Note that $Y$, $\boldsymbol{X}$ and $\boldsymbol{X}^*$ have been scaled by their respective standard deviations. 
$S \in \{0,1\}$ denote membership in the experimental sample and target sample from the respective populations.
Let $g(.)$ and $h(.)$ denote the conditional mean for the outcome and selection models above, and let $\Phi_\text{out}(Y,A,\boldsymbol{X};\beta)$ and $\Phi_\text{sel}(S,\boldsymbol{X};\gamma)$ denote the outcome and selection loss functions, respectively. Here, the outcome model is using a linear regression model and the selection model is a logistic regression model. In this case, $\Phi_\text{out}$ is the squared error loss, $(Y-g(A,\boldsymbol{X}))^2$, and $\Phi_\text{sel}$ is proportional to the binomial negative log-likelihood, $S \times \log (1+e^{-2h(\boldsymbol{X}^*)})-(1-S)\times \log (1+e^{-2h(\boldsymbol{X})})$. 
We define $2p+2$ groups of vector $\boldsymbol{\alpha} = (\beta, \gamma)$ such that
$\alpha_1=(\beta_1), ..., \alpha_p=(\beta_p), \alpha_{p+1}=(\beta_{p+1}, \gamma_1),..., \alpha_{2p}=(\beta_{2p}, \gamma_p), \alpha_{2p+1}=(\beta_{2p+1}, \gamma_{p+1}), \alpha_{2p+2}=(\beta_{2p+2})$. These groups consists of $p$ outcome main effect coefficients, $p$ coefficients of outcome treatment interaction grouped with the corresponding coefficients from the selection model, a group of intercepts from both models, and finally the treatment coefficient as its own group. These coefficients groupings across the selection and outcome models forces covariates to enter and leave the final model simultaneously. The intent behind these groupings is that the likelihood that coefficients that are weak treatment interactions are included in the model and can be boosted by considering the corresponding coefficients from the selection model. 

This simultaneous variable selection procedure then consists of solving: 
\begin{align}
\label{alpha.eq}
    \hat\alpha(\lambda) = &\underset{\boldsymbol{\alpha}}{argmin} ( \frac{1}{n} \sum^n_{i=1} \Phi_\text{out}(Y,A,\boldsymbol{X}_i;\beta)  \\
    &+ \frac{1}{n+m} 
    \big( 
    \sum^{n}_{i=1}\Phi_\text{sel}(S=1,\boldsymbol{X}_i;\gamma)
    +
    \sum^{m}_{i=1}\Phi_\text{sel}(S=0,\boldsymbol{X}_i^*;\gamma)
    \big)
    + \lambda \sum^{2p+2}_{k=1}w_k ||\alpha_k||_2 \nonumber
\end{align}
We emphasize that the outcome coefficients are estimated only using the experimental sample and the selection coefficients are estimated using both the experimental and target samples. 

In the group lasso \cite{Yuan}, a common choice for the penalty weight for group $k$ is $w_k=\sqrt{c_k}$, where $c_k$ is the cardinality (i.e., the number of elements in the group) of group $k$.
Here, however, the weights are set as:
\begin{align}
w_k &= \begin{cases}
\frac{1}{|v_k|}, & \text{for } k=1,...,p\\
\frac{\sqrt{2}}{|v_k|}, & \text{for } k=p+1,...,2p\\
0, & \text{for } k=2p+1, 2p+2
\end{cases}
\end{align}

For $k=1,...,2p$, the numerator corresponds to the cardinality of group $k$, but the denominator $v_k \neq 0$ is an estimate of the regression coefficient for the treatment interaction of covariate $k$ in the ``full" outcome model that includes the main effects and treatment interactions with no variable selection. This choice in denominator strongly penalizes covariates that are not treatment interactions (i.e., a small $|v_k|$) associated with the outcome even if they are strongly associated with the selection. For the intercepts $k=2p+1, 2p+2$, we set the weights as $w_k = 0$ so that they are not penalized. The goal of the proposed weights used here is to select covariates that are either strong treatment interactions associated with the outcome, or strongly associated with the selection and a treatment interaction for the outcome. 

A modified version of generalized cross-validation (GCV) is used to solve for $\hat\alpha(\lambda)$ because we consider both outcome and selection loss functions, but want to apply GCV to only the outcome model. This is the same GCV used by GLiDeR \cite{Koch}. In the general group lasso, the GCV at a particular value of $\lambda$ is $\frac{RSS}{(1-df/n)^2}$, where RSS is the residual sum of squares and $df=\sum_{k=1}^K I(||\hat\xi_k||>0) + \sum_{k=1}^K \frac{||\hat\xi_k||}{||\tilde\xi_k||}(d_k-1)$ where $\hat\xi_k$ and $\tilde\xi_k$ are respectively the adaptive group lasso and least squares estimators of the $k=1,...,K$ group of coefficients, each with group size $d_k$.
Here, since we want a model selection procedure for only the outcome, both the numerator and the denominator only use the parts of $\hat\alpha(\lambda)$ that correspond to the coefficients of the outcome model:
\begin{align}
    GCV(\lambda) &= \frac{\sum_{i=1}^n (
    Y_i - \hat\beta_{2p+1}(\lambda) - \hat\beta_{2p+2}(\lambda)A_i
    - \sum_{k=1}^{p} \hat\beta_k(\lambda)X_{ki} +\hat\beta_{p+k}(\lambda)X_{ki}A_i)^2}
    {(1-(2+\sum_{k=1}^{2p} I(||\hat\beta_k(\lambda)||>0) + \sum_{k=1}^{2p} \frac{|\hat\beta_k(\lambda)|}{|v_k|})/n)^2}
\end{align}

Given a solution $\hat\alpha(\lambda)  = \{\hat\beta(\lambda), \hat\gamma(\lambda)\}$, we can plug $\hat\beta(\lambda)$ into the outcome model to obtain the estimated TATE: 
\begin{align}
\label{delta.eq}
    \Delta^* &= \hat\beta(\lambda)_{2p+2}+\frac{1}{m}\sum_{i}\sum_{j}\hat\beta(\lambda)_{p+j}X^*_{ij}
\end{align}
The confidence intervals can be calculated with bootstrapping. Since TATE estimates from the GLAVeS simulations showed significant bias despite good variable selection accuracy, we also introduce OLSGLAVeS, where variable selection for generalization is performed using GLAVeS and then the TATE estimate is obtained using OLS estimates for the reduced model.

\subsection{Survey Weights}
Often in public health, our target population is a broad population (e.g., the overall U.S. adult population) for which it is difficult to obtain a representative sample via random sampling. However, there are many large surveys that can provide a wealth of information on the target population. 
It is important to note that these raw analytical survey datasets are not representative samples of the target population; they are the result of complex survey designs that over-sample or under-sample certain demographic groups, adjust for non-response, etc. 
In order to calibrate from these survey samples to the target population of interest, survey weights are typically included in the final analytic datasets and used to obtain unbiased estimators of the target population parameters. 
Survey weights can easily be incorporated into GLAVeS with the additional changes outlined below.

Let $r_i$ denote the raw survey weights from the sample of the target population with $i=1,...,m$ subjects. In GLAVeS, we define a weight $\Omega_i$ such that subjects from the experimental sample are given a weight of 1, and subjects from the target population are given survey weights standardized to have a mean of 1: 
\begin{align}
\Omega_i &= \frac{m}{\sum_i r_i} r_i, & \text{for } i=1,...,m
%\begin{cases}
%1, & \text{for } i=1,...,n \\
%\frac{n+m}{\sum_i r_i}r_i, & \text{for } i=n+1,...,n+m \\
%\end{cases}
\end{align}
This weight can then be included in equations \ref{alpha.eq} as: 
\begin{align}
    \hat\alpha(\lambda) &= \underset{\boldsymbol{\alpha}}{argmin} ( \frac{1}{n} \sum^n_{i=1} \Phi_\text{out}(Y,A,\boldsymbol{X};\beta) \\
    & + \frac{1}{n+m} \big(   
    \sum^{n}_{i=1} \Phi_\text{sel}(S=1,\boldsymbol{X}_i;\gamma) 
    + \sum^{m}_{i=1} \Omega_i \Phi_\text{sel}(S=0,\boldsymbol{X}^*_i;\gamma) 
    \big) 
    + \lambda \sum^{2p+2}_{k=1}w_k ||\alpha_k||_2 \nonumber
\end{align}
where the survey weights are only incorporated into the selection model, and included in equation \ref{delta.eq} for the final TATE: 
\begin{align}
    \hat{\Delta}^* &= \hat\beta(\lambda)_{2p+2}+\frac{1}{m}\sum_{i}\sum_{j}\Omega_i\hat\beta(\lambda)_{p+j}X^*_{ij}
\end{align}

%%%%%%%%%%%%%%% SIMULATIONS %%%%%%%%%%%%%%%%%%%%
\section{Simulations}
\label{Gsims}
We completed a simulation study to investigate the statistical properties of the variable selection method described in section \ref{glid}. We looked at two overarching problem settings in our simulations: one with 8 total covariates and another with 15 total covariates. 
In each of these settings, we varied the size of the coefficients for the treatment interactions and the correlation among covariates. A complete description of our simulation study can be found below. 

\subsection{Data generation}
We considered a range of 8 different scenarios with varying number of covariates, correlation, treatment interaction strengths, and level of dissimilarity between the experimental sample and target population sets. 
For each replicate experimental sample set, we simulated 600 individuals, indexed by $i$, with 8 total covariates for scenarios 1-4, and 15 total covariates in scenarios 5-8. 
The covariates $\boldsymbol{X}_i=(X_{1i}, X_{2i}, ..., X_{pi})$ are generated from a multivariate standard Gaussian distribution if there is no correlation (scenarios 1,2,5 and 6).
When there is correlation specified for a scenario, $\boldsymbol{X}_i=(X_{1i}, X_{2i}, ..., X_{pi})$ are generated from a multivariate Gaussian distribution $N(0,\Sigma)$, where $\Sigma$ is a matrix with diagonal element equal to 1 (i.e., a correlation matrix), such that each correlated covariate set specified in Table \ref{scenarios} has a compound symmetric covariance structure with a correlation of 0.6.
We also generated a binary treatment variable for each individual, $A_i$ from a Bernoulli distribution with probability 0.5. The outcome for each individual, $Y_i$, is generated from $N(A_i + g(\boldsymbol{X}_i), 1)$ where $g(.)$ is the outcome model specified in Table \ref{scenarios}. 

\begin{table}[!ht]
\small
    \centering
    \begin{threeparttable}
    \begin{tabular}{cclll}
    \toprule
      Scenario & $p$  &  Outcome Model $g(.)$ & 
      \vtop{\hbox{\strut Correlated}\hbox{\strut Covariates}}
      & \vtop{\hbox{\strut Covariates Distributed}\hbox{\strut $N(1, 1)$ in Target Set}}     \\
    \midrule
      1 & 8 & $X_1+X_3+0.1AX_1+0.1AX_3$ &  &$X_1, X_2, X_3$  \\
      2 & 8 & $X_1+X_3+0.05AX_1+0.05AX_3$ &  & $X_1, X_2, X_3$   \\
      3 & 8 & $X_1+X_3+0.1AX_1+0.1AX_3$ & $\{X_1, X_2\}, \{X_3, X_4\}$ & $X_1, X_2, X_3$ \\
      4 & 8 & $X_1+X_3+0.05AX_1+0.05AX_3$ & $\{X_1, X_2\}, \{X_3, X_4\}$ & $X_1, X_2, X_3$ \\
      5 & 15 & $X_1+X_2+0.1AX_1+0.1AX_3+0.1AX_5$ &  &$X_1, ..., X_7$  \\
      6 & 15 & $X_1+X_2+0.05AX_1+0.05AX_3+0.05AX_5$ & & $X_1, ..., X_7$  \\
      7 & 15 & $X_1+X_2+0.1AX_1+0.1AX_3+0.1AX_5$
      & \vtop{\hbox{\strut $\{X_1, X_2\}, \{X_3, X_4\},$}
      \hbox{\strut $\{X_5, X_6, X_7\},$}
      \hbox{\strut $\{X_8, X_9, X_{10}\}$}
      }
      & $X_1, ..., X_7$ \\
      8 & 15 & $X_1+X_2+0.05AX_1+0.05AX_3+0.05AX_5$
      & \vtop{\hbox{\strut $\{X_1, X_2\}, \{X_3, X_4\},$} 
      \hbox{\strut $\{X_5, X_6, X_7\},$}
      \hbox{\strut $\{X_8, X_9, X_{10}\}$}
      }
      & $X_1, ..., X_7$\\
     \bottomrule
    \end{tabular}
    \end{threeparttable}
    \caption{The 8 scenarios considered. }
    \label{scenarios}
\end{table}

For each replicate target set, we simulated only the covariates, $\boldsymbol{X}^*_i=(X^*_{1i}, X^*_{2i}, ..., X^*_{pi})$, for 300 individuals.  These covariates are again generated from a multivariate standard Gaussian distribution unless otherwise indicated by the different scenarios in Table \ref{scenarios}. 
%The data generation of scenario 3 is provided in Figure \ref{exDG} as an example. 
%Onenote Glider, Data Generation page for code

\subsection{Methods Implemented}
We implemented 6 different methods, along with GLAVeS as outlined in the methods section. %\subsubsection{Simple difference} 
The first method we implemented is a naive method that takes the simple difference between the average outcome in the experimental sample under treatment and under control: 
\begin{equation*}
        \Delta_{\text{S.Diff}} = \frac{\sum_{i=1}^nY_iI(A_i=1)}{\sum_{i=1}^nI(A_i=1)} - \frac{\sum_{i=1}^nY_iI(A_i=0)}{\sum_{i=1}^nI(A_i=0)}
\end{equation*}
It disregards any distributional differences between the experimental sample and target population and gives an indication of what would happen if no adjustment was made to account for the different populations. This method is denoted as ``S.Diff" in the simulation results below. 
The next set of methods use different variable selection models to fit the outcome models. All use the plug-in estimator for the TATE from Equation \ref{tate.eq}.

%\subsubsection{OLS counterfactual models} 
The second method fits separate linear regression models for the treatment and control arms in the experimental sample with all the covariates. No variable selection procedure is used. This method is denoted as ``OLS" in the simulation results:
\begin{equation*}
        \hat{\Delta}_{\text{OLS}} = \frac{1}{m}\sum^m_{i=1}
        \boldsymbol{X}_i^{*T} \hat{\boldsymbol{\beta}}_{\text{OLS,}A=1} - 
        \boldsymbol{X}_i^{*T} \hat{\boldsymbol{\beta}}_{\text{OLS,}A=0}
\end{equation*}

%\subsubsection{LASSO with interactions} 
The third method fits a linear model with all the covariates and possible treatment interactions with a lasso (i.e., L1) penalty using the experimental sample. We use cross-validation to select the penalty parameter $\lambda$ that minimizes mean squared prediction error. We implemented this using the \texttt{glmnet} package in R \cite{glmnet}. This method is denoted as ``IntLASSO" in the simulation results:
%\begin{equation*}
%        \Delta_{\text{IntLASSO}} =  \begin{bmatrix} \boldsymbol{1}_{m\times 1} & \boldsymbol{X}^* & \boldsymbol{X}^* \end{bmatrix} \boldsymbol{\beta}_{\text{IL}} -
%        \begin{bmatrix} \boldsymbol{0}_{m\times 1} & \boldsymbol{X}^* & \boldsymbol{0}_{m\times p} \end{bmatrix} \boldsymbol{\beta}_{\text{IL}}
%\end{equation*}

%\subsubsection{LASSO counterfactual models} 
The fourth method fits separate LASSO models for the treatment and control arms in the experimental sample using all covariates. We again use cross-validation to select the penalty parameter $\lambda$ that minimizes mean squared error in each model. We implemented this using the \texttt{glmnet} package in R \cite{glmnet}.  This method is denoted as ``LASSO" in the simulation results below: 
%\begin{equation*}
%        \Delta_{\text{LASSO}} = \boldsymbol{X}^* \boldsymbol{\beta}_{\text{LASSO,}A=1} -
%        \boldsymbol{X}^* \boldsymbol{\beta}_{\text{LASSO,}A=0}
%\end{equation*}

%\subsubsection{GLAVeS and OLSGLAVeS} 
The final methods we implement are GLAVeS as described in Section \ref{Gmethods} and ``OLSGLAVeS", where GLAVeS is used solely as a variable selection method and then an ordinary least squares regression model is fitted using the selected covariates and treatment interactions. This linear regression model is then applied to the target population to obtain the TATE estimate.  

%%%%%%%%%%%%%%% SIMULATION RESULTS %%%%%%%%%%%%%%%%%%%%

\subsection{Simulation Results}
\label{sbsimres}

For each scenario, we calculated the bias and MSE of each method averaged over 1000 simulations. The biases of the various methods are shown in Table \ref{biastable}. S.Diff, where no adjustments were made to account for the different populations, always yields the largest absolute bias across all the scenarios considered. Out of the methods that perform variable selection, OLSGLAVeS has the lowest absolute bias for all scenarios tested. Although OLS has the lowest bias for scenarios 5 through 8 where there were 15 covariates, it does not perform variable selection. Overall, the OLSGLAVeS estimates are only slightly more biased than the OLS estimates, where no variable selection is performed and a low bias was expected. 

\begin{table}[ht]
\centering
\begin{tabular}{crrrrrr}
  \hline
Scenario & S.Diff & OLS & IntLASSO & LASSO & GLAVeS & OLSGLAVeS \\ 
  \hline
1 & -0.202 & 0.003 & -0.103 & 0.003 & 0.123 & \bf{-0.001} \\ 
  2 & -0.102 & 0.003 & -0.079 & 0.003 & 0.136 & \bf{0.001} \\ 
  3 & -0.196 & 0.006 & -0.094 & 0.006 & 0.129 & \bf{0.003} \\ 
  4 & -0.096 & 0.006 & -0.072 & 0.006 & 0.146 & \bf{0.003} \\ 
  5 & -0.304 & 0.000 & -0.198 & -0.066 & 0.118 & \bf{-0.017} \\ 
  6 & -0.154 & 0.000 & -0.151 & -0.050 & 0.134 & \bf{-0.015} \\ 
  7 & -0.307 & 0.001 & -0.192 & -0.064 & 0.069 & \bf{-0.018} \\ 
  8 & -0.157 & 0.001 & -0.153 & -0.049 & 0.090 & \bf{-0.016} \\ 
   \hline
\end{tabular}
\caption{Biases for scenarios 1-8 outlined in Table \ref{scenarios}}
\label{biastable}
\end{table}

The MSEs of the various methods are shown in Table \ref{msetable}. In the scenarios with 8 total covariates, OLSGLAVeS has the lowest MSEs when the treatment interactions have a true coefficient of 0.1 (scenarios 1 \& 3) and IntLASSO has the lowest MSEs when the treatment interactions have a true coefficient of 0.05 (scenarios 2 \& 4). S.Diff always yields the highest MSE across all the scenarios considered.
In the scenarios with 15 covariates, LASSO has the lowest MSE for all scenarios except when the covariates are correlated and has a true treatment interaction coefficient of 0.1 (scenario 7), where GLAVeS has the lowest MSE. Across all the scenarios, LASSO and OLSGLAVeS consistently have the lowest MSE: they have either the lowest MSE or close to the lowest MSE across all methods and for all scenarios. 
Additional simulation scenarios with results for the bias and MSE can be found in Appendix \ref{G_app}.

\begin{table}[ht]
\centering
\begin{tabular}{crrrrrr}
  \hline
Scenario & S.Diff & OLS & IntLASSO & LASSO & GLAVeS & OLSGLAVeS\\ 
  \hline
  1 & 0.062 & 0.027 & 0.030 & 0.025 & 0.036 & \bf{0.024} \\ 
  2 & 0.031 & 0.027 & \bf{0.022} & 0.025 & 0.043 & 0.026 \\ 
  3 & 0.061 & 0.025 & 0.028 & 0.025 & 0.038 & \bf{0.024} \\ 
  4 & 0.031 & 0.025 & \bf{0.021} & 0.025 & 0.042 & 0.023 \\ 
  5 & 0.112 & 0.061 & 0.063 & \bf{0.044} & 0.055 & 0.052 \\ 
  6 & 0.043 & 0.061 & 0.040 & \bf{0.037} & 0.055 & 0.047 \\ 
  7 & 0.125 & 0.036 & 0.061 & 0.037 & \bf{0.032} & 0.036 \\ 
  8 & 0.055 & 0.036 & 0.041 & \bf{0.029} & 0.034 & 0.033 \\ 
   \hline
\end{tabular}
\caption{MSEs for scenarios 1-8 outlined in Table \ref{scenarios}}
\label{msetable}
\end{table}

For each of the scenarios from Table \ref{scenarios}, we also output the average sensitivity and specificity of each method for identifying the correct treatment interactions over 1000 simulations. For S.Diff, since no adjustment is performed and no treatment interactions are used, the sensitivity will always be 0 and the specificity will always be 1. Conversely, for OLS, all possible treatment interactions are included in the model, so the sensitivity will always be 1 and the specificity will always be 0. A treatment interaction is considered ``selected" by the LASSO model if it is picked up by either the treatment or control models. Since OLSGLAVeS uses the variables selected by GLAVeS, they will always have the same sensitivity and specificity. These sensitivity and specificity measures are shown for scenario 1- 8, along with the MSE, bias, and standard deviations from 1000 simulations in Tables \ref{S1.4} and \ref{S5.8}.

\begin{table}[!htbp]
\centering
\subfloat[Scenario 1]{ %cond 17
\begin{tabular}{r|lllll}
  \hline
  & MSE & Bias & SD & SE & SP \\ 
  \hline
  S.Diff & 0.062 & -0.202 & 0.145 & 0 & 1 \\ 
  IntLASSO & 0.03 & -0.103 & 0.141 & 0.8 & 0.67 \\ 
  LASSO & 0.025 & 0.003 & 0.159 & 1 & 0.38 \\ 
  GLAVeS & 0.036 & 0.123 & 0.146 & 0.95 & 0.78 \\ 
  OLSGLAVeS & 0.024 & -0.001 & 0.156 & 0.95 & 0.78 \\ 
  OLS & 0.027 & 0.003 & 0.166 & 1 & 0 \\ 
   \hline
\end{tabular}
}
\quad
\subfloat[Scenario 2]{ %cond 18
\begin{tabular}{rlllll}
  \hline
 & MSE & Bias & SD & SE & SP \\ 
  \hline
& 0.031 & -0.102 & 0.143 & 0 & 1 \\ 
& 0.022 & -0.079 & 0.127 & 0.6 & 0.68 \\ 
& 0.025 & 0.003 & 0.159 & 1 & 0.38 \\ 
& 0.043 & 0.136 & 0.158 & 0.9 & 0.74 \\ 
& 0.026 & 0.001 & 0.161 & 0.9 & 0.74 \\ 
& 0.027 & 0.003 & 0.165 & 1 & 0 \\ 
   \hline
\end{tabular}
}
\par\medskip
%%%
\subfloat[Scenario 3]{ %cond 21
\begin{tabular}{r|lllll}
  \hline
 & MSE & Bias & SD & SE & SP \\ 
  \hline
  S.Diff & 0.061 & -0.196 & 0.15 & 0 & 1 \\ 
  IntLASSO & 0.028 & -0.094 & 0.139 & 0.8 & 0.66 \\ 
  LASSO & 0.025 & 0.006 & 0.157 & 1 & 0.36 \\ 
  GLAVeS & 0.038 & 0.129 & 0.145 & 0.93 & 0.71 \\ 
  OLSGLAVeS & 0.024 & 0.003 & 0.154 & 0.93 & 0.71 \\ 
  OLS & 0.025 & 0.006 & 0.158 & 1 & 0 \\ 
   \hline
\end{tabular}
}
\quad
\subfloat[Scenario 4]{ %cond 22
\begin{tabular}{rlllll}
  \hline
 & MSE & Bias & SD & SE & SP \\ 
  \hline
& 0.031 & -0.096 & 0.147 & 0 & 1 \\ 
& 0.021 & -0.072 & 0.125 & 0.59 & 0.68 \\ 
& 0.025 & 0.006 & 0.157 & 1 & 0.35 \\ 
& 0.042 & 0.146 & 0.145 & 0.9 & 0.68 \\ 
& 0.023 & 0.003 & 0.152 & 0.9 & 0.68 \\ 
& 0.025 & 0.006 & 0.158 & 1 & 0 \\ 
   \hline
\end{tabular}
}
%%%
\caption{MSE, bias, standard deviations (SD), sensitivity (SE) and specificity (SP) of various methods for scenarios 1 - 4 where there were 8 total covariates.}
\label{S1.4}
\end{table}

In the 8 total covariates scenarios shown in Table \ref{S1.4}, GLAVeS/OLSGLAVeS have the best sensitivity and specificity balance, with a sensitivity of 0.9 to 0.95 and specificity of 0.68 to 0.78. While the sensitivity of LASSO is 1 in all scenarios, its specificity is much lower at 0.35 to 0.38, meaning it is including too many unrelated variables along with the true treatment interactions. 
Both the sensitivity and specificity of IntLASSO are lower than that of GLAVeS/OLSGLAVeS in all these scenarios. 

\begin{table}[!htbp]
\centering
\subfloat[Scenario 5]{ %cond 25
\begin{tabular}{r|lllll}
  \hline
  & MSE & Bias & SD & SE & SP \\ 
  \hline
  S.Diff & 0.112 & -0.304 & 0.143 & 0 & 1 \\ 
  IntLASSO & 0.063 & -0.198 & 0.154 & 0.7 & 0.73 \\ 
  LASSO & 0.044 & -0.066 & 0.2 & 0.88 & 0.43 \\ 
  GLAVeS & 0.055 & 0.118 & 0.203 & 0.75 & 0.75 \\ 
  OLSGLAVeS & 0.052 & -0.017 & 0.228 & 0.75 & 0.75 \\ 
  OLS & 0.061 & 0 & 0.248 & 1 & 0 \\ 
   \hline
\end{tabular}
}
\quad
\subfloat[Scenario 6]{ %cond 26
\begin{tabular}{rlllll}
  \hline
 & MSE & Bias & SD & SE & SP \\ 
  \hline
& 0.043 & -0.154 & 0.141 & 0 & 1 \\ 
& 0.04 & -0.151 & 0.129 & 0.44 & 0.76 \\ 
& 0.037 & -0.05 & 0.185 & 0.73 & 0.47 \\ 
& 0.055 & 0.134 & 0.193 & 0.6 & 0.74 \\ 
& 0.047 & -0.015 & 0.216 & 0.6 & 0.74 \\ 
& 0.061 & 0 & 0.248 & 1 & 0 \\ 
   \hline
\end{tabular}
}
\par\medskip
%%%
\subfloat[Scenario 7]{ %cond 29
\begin{tabular}{r|lllll}
  \hline
 & MSE & Bias & SD & SE & SP \\ 
  \hline
  S.Diff & 0.125 & -0.307 & 0.176 & 0 & 1 \\ 
  IntLASSO & 0.061 & -0.192 & 0.156 & 0.64 & 0.75 \\ 
  LASSO & 0.037 & -0.064 & 0.181 & 0.82 & 0.51 \\ 
  GLAVeS & 0.032 & 0.069 & 0.166 & 0.7 & 0.75 \\ 
  OLSGLAVeS & 0.036 & -0.018 & 0.19 & 0.7 & 0.75 \\ 
  OLS & 0.036 & 0.001 & 0.19 & 1 & 0 \\ 
   \hline
\end{tabular}
}
\quad
\subfloat[Scenario 8]{ %cond 30
\begin{tabular}{rlllll}
  \hline
 & MSE & Bias & SD & SE & SP \\ 
  \hline
& 0.055 & -0.157 & 0.175 & 0 & 1 \\ 
& 0.041 & -0.153 & 0.131 & 0.37 & 0.79 \\ 
& 0.029 & -0.049 & 0.165 & 0.66 & 0.55 \\ 
& 0.034 & 0.09 & 0.162 & 0.57 & 0.76 \\ 
& 0.033 & -0.016 & 0.18 & 0.57 & 0.76 \\ 
& 0.036 & 0.001 & 0.19 & 1 & 0 \\ 
   \hline
\end{tabular}
}
%%%
\caption{MSE, bias, standard deviations (SD), sensitivity (SE) and specificity (SP) of various methods for scenarios 5 - 8 where there were 15 total covariates.}
\label{S5.8}
\end{table}

In the 15 total covariates scenarios shown in Table \ref{S5.8}, GLAVeS/OLSGLAVeS have the best sensitivity and specificity balance again, with a sensitivity of 0.57 to 0.75 and specificity of 0.74 to 0.76. LASSO is again including too many unrelated variables with a specificity of 0.43 to 0.55 even though it is catching more true treatment interactions with a sensitivity of 0.66 to 0.88. IntLASSO's sensitivity and specificity are either similar or lower than GLAVeS/OLSGLAVeS in all scenarios. 

tobacco (typical of commercial brands) to 0.4 mg/g, and the primary outcome was the number of cigarettes smoked per day during week 6. At six weeks, smokers randomized to the 0.4 mg/g group smoked significantly fewer cigarettes than smokers randomized to the 15.5 mg/g group \cite{Donny}.
CENIC Project 2 was a double-blind, randomized, parallel-design study to evaluate the impact of immediate vs. gradual reduction in nicotine content conducted at 10 US sites between July 2014 and March 2017. A volunteer sample of daily smokers with no intention to quit within 30 days completed a 2 week baseline smoking period, after which they were randomly assigned to 1 of 3 treatment conditions:
1. Immediate nicotine reduction, where participants were assigned to use cigarettes with a nicotine content of 0.4 mg/g tobacco. 
2. Gradual nicotine reduction, where participants smoked progressively lower nicotine content cigarettes for a period of one month each until receiving the same 0.4 mg/g tobacco VLNC cigarettes as the immediate reduction group for the last month of the study. 
3. Usual nicotine control, where participants were given their assigned cigarettes with a nicotine content of 15.5 mg/g tobacco, approximately equivalent to the average nicotine content of commercial cigarettes. 
Participants were given study cigarettes for a period of 20 weeks and outcome measures assessed included biomarkers of smoke exposure, and participants' measures of smoking behavior were recorded. The dataset contains 46 demographic and baseline covariates such as age, gender, race, education, cigarettes per day, total nicotine equivalents (TNE, a common measure of internal nicotine dose which includes nicotine and its metabolites), carbon monoxide levels, etc. Smokers randomized to immediate reduction in nicotine content had significantly reduced biomarkers of tobacco exposure over time compared to somkers randomized to gradual reduction or the control group \cite{Hatsukami}.

In the context of tobacco regulatory science, the target population is the entire United States smoking population. Systematic differences between the RCT population and the target population are not a problem in the absence of treatment effect heterogeneity (i.e., when the effect of the intervention varies as a function of participant characterization), but, in the presence of treatment effect heterogeneity, the treatment effect estimated in the RCT may be biased relative to the effect in the target population. Concerns about this bias in the context of tobacco regulatory science are particularly acute because any regulation will impact all US smokers, not just those that participate in the intervention.
By generalizing the treatment effects of VLNCs in the CENIC studies over to the population of the Population Assessment of Tobacco Use and Health (PATH) study, we would be able to estimate the effect of a potential nicotine reduction policy on the overall US smoking population.

We generalize the effect of potential tobacco product regulations evaluated in the context of RCTs to the entire US smoking population using the PATH, a prospective cohort study of tobacco-use patterns, risk perceptions, and attitudes towards current tobacco products among tobacco users and non-users in the United States \cite{path}. This longitudinal study, supported by the FDA CTP and the National Institutes of Health, is funded to collect data annually through 2024. Baseline data was collected during September 2013 and December 2014 and included 45,971 adults and youths \cite{path}. Many of the survey and biomarker covariates collected in this study correspond to identical or analogous covariates in the CENIC studies. 

\begin{table}[!htbp]
\centering
\begin{tabular}{lcc}
\hline
\textbf{Variable} & \multicolumn{2}{c}{\textbf{Median[IQR] or Count(\%)}} \\
\hline
& \bf \underline{CENIC} & \bf \underline{PATH} \\ 
  n & 538 & 8351 \\ 
  Age & 48 [35, 56] & 38 [26, 52] \\ 
  Gender = Male & 290 (53.9) & 4333 (51.9) \\ 
  Baseline Weight (kg) & 83 [70, 96] & 77 [66, 91] \\ 
  Race &  &  \\ 
  \qquad    White & 336 (62.5) & 6375 (76.3) \\ 
  \qquad    Black & 164 (30.5) & 1215 (14.5) \\ 
  \qquad    Other & 38 (7) & 761 (9.1) \\ 
  Education &  &  \\ 
  \qquad    HS or less & 49 (9.1) & 2490 (29.8) \\ 
  \qquad    HS Grad & 170 (31.6) & 2202 (26.4) \\ 
  \qquad    College or more & 319 (59.3) & 3659 (43.8) \\ 
  Use of Menthol Cigarettes = Menthol & 243 (45.2) & 3256 (39) \\ 
  Baseline Total Cigarettes/Day & 16 [11, 22] & 10 [5, 15] \\
\hline
\end{tabular}
\caption{Demographic and baseline characteristics of CENIC and Wave 1 PATH participants.}
\label{tab1}
\end{table}

\begin{figure}[!htbp]
\centering
\includegraphics[width=0.75\textwidth]{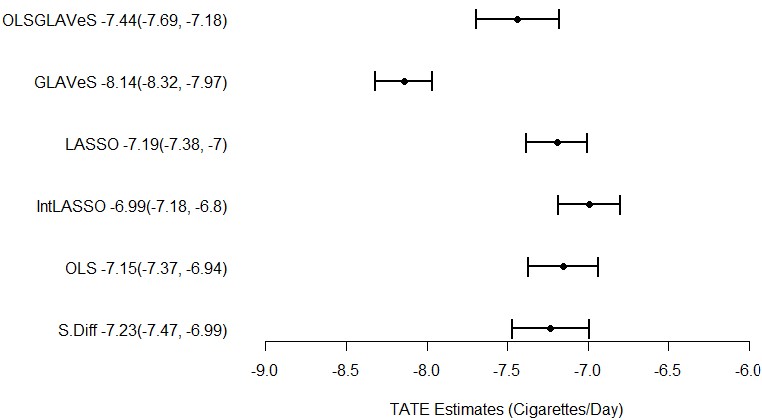}
\caption{Forest plot of point estimates and corresponding bootstrapped percentile 95\% confidence intervals of the TATE estimates of VLNC cigarettes generalized to the overall U.S. smoking population using various estimation methods. }
\label{tates}
\end{figure}

The summary of covariates that were available in both datasets are presented in Table \ref{tab1}. Overall, the Wave 1 PATH population is younger, lighter, more white, less educated, uses less menthol cigarettes, and smoked less at baseline compared to the CENIC population. The SATE of VLNC cigarettes in the CENIC population is estimated to be -7.65 cigarettes/day, meaning smokers using VLNC cigarettes in the CENIC population will smoke 7.65 cigarettes/day less on average than smokers using the control cigarettes. The PATH dataset contains survey weights that we used to generalize the different TATE estimates to the overall U.S. smoking population. 
The TATE estimates following treatment effect generalization is computed using all the methods described in Section \ref{Gsims} and the point estimates and corresponding percentile 95\% confidence intervals of 100 bootstrapped samples are presented in Figure \ref{tates}. A comparison of the selected covariates and their coefficients are presented in Table \ref{vscoefs}.

From the TATE estimates of the various methods, we see that most estimates place the treatment effect of VLNC cigarettes to be between 7 to 7.5 cigarettes/day when generalized from CENIC to the overall U.S. smoking population. Only GLAVeS gives an estimate outside that range of a 8.14 cigarettes/day difference. 
%don't really know how to relate to sim results?

In terms of variable selection for generalization, GLAVeS/OLSGLAVeS was the most selective, picking up only age as a treatment interaction after penalizing for both the relation to outcome and distributional difference of covariates between the CENIC and PATH populations. IntLASSO included every possible treatment interaction, and LASSO only excluded race as an interaction. Since S.Diff does not adjust for the distributional differences between the populations, no treatment interactions are included. OLS does not perform any variable selection, and all possible treatment interactions are included in the linear regression model. 

\begin{table}[!htbp]
\centering
\begin{tabular}{lcccccc}
\hline
\textbf{Variable} & \textbf{S.Diff} &\textbf{IntLASSO} & \textbf{LASSO} & \textbf{GLAVeS} & \textbf{OLSGLAVeS} & \textbf{OLS}\\
\hline
  Age && 0.11 & 0.12 & 0.14 & 0.17 & 0.14\\
  Gender = Male && -0.12 & -1.36 & & & -0.15\\
  Baseline Weight (kg) && -0.04 & -0.01 & & & -0.05\\
  Race = White && 0.87 & & & & 2.26\\
  Race = Black && 0.55 & & & & 1.42\\
  Education = HS or less && 0.28 & -0.18 & & & 0.33\\
  Education = HS Grad && -0.08 & -0.62 & & & -0.27\\
  \vtop{\hbox{\strut Use of Menthol Cigarettes}\hbox{\strut \qquad = Menthol }} && 2.38 & 0.72 & & &3.15 \\
  \vtop{\hbox{\strut Baseline Total}\hbox{\strut \qquad Cigarettes/Day}}   && -0.15 & -0.22 & & & -0.17 \\
\hline
\end{tabular}
\caption{Variables selected and estimated coefficients of treatment interaction terms for various estimation methods.}
\label{vscoefs}
\end{table}

%%%%%%%%%%%%%%% DISCUSSION  %%%%%%%%%%%%%%%%%%%%
\section{Discussion}
\label{disc}

%how method fit into lit
In the current causal inference literature, there are many data-driven variable selection methods for potential confounders. However, variable selection for the purpose of treatment effect generalization is a poorly explored problem with no established solutions, despite its importance for many public health problems. %In this paper, we use a simulation study to first clarify the issues surrounding variable selection in the context of treatment effect generalization, and then propose 
Standard variable selection methods are sufficient to capture strong treatment interactions, but weaker interactions are often missed, with significant implications for the accuracy of TATE estimations in the case of differentially distributed covariates. 
GLAVeS is designed to catch these weaker treatment interactions by supplementing the penalization of the treatment interaction terms with the probability of selection into the target population using a modified adaptive group lasso approach.

%summary of results
In all the simulation scenarios we explored, GLAVeS is able to pick out the treatment interactions with a higher combined sensitivity and specificity than any other methods we implemented. This accuracy in identifying the correct treatment interactions is reflected in the low bias of the OLSGLAVeS TATE estimates. That said, the low bias of the OLSGLAVeS estimates do not necessarily translate to the minimum MSE because the standard deviation of the estimates are higher compared to the other methods we implemented. However, we argue that the bias of the TATE estimate is a more important metric since having a high bias would lead to poor coverage on the 95\% confidence intervals for the estimates. 
Overall, in treatment effect generalization applications where there are potentially many weak treatment interactions, we recommend the use of GLAVeS to select the covariates that should be used in an ordinary least squares regression model to obtain an unbiased estimation of the TATE. 

%talk about PATH application?

%what we could do better in the future
In the future, it would be interesting to repeat the simulations under different settings, such as decreasing the number of individuals in the experimental sample to explore the performance of various methods when the generalization is based on an experimental sample smaller than 600. 
Another possible extension would be to modify other methods for confounder variable selection besides GLiDeR to fit the treatment effect generalization framework, and to test the performance of those methods against GLAVeS. 

\section*{Acknowledgements}

The author(s) disclosed receipt of the following financial support
for the research, authorship, and/or publication of this
article: This study was funded by the
National Cancer Institute (award numbers R01CA214825
and R01CA225190), the National Institute on Drug Abuse
(award numbers R01DA046320, R03DA041870, and U54-
DA031659) and National Center for Advancing Translational
Science (award number UL1TR002494). The content is solely
the responsibility of the authors and does not necessarily represent
the official views of the National Institutes of Health
or Food and Drug Administration.

% REFERENCES
\newpage
\bibliographystyle{bibft}\it
\bibliography{bibfile}

% \appendix

\end{document}